%% file: fast_qrs.tex
\documentclass[english,12pt]{article}
\usepackage[T1]{fontenc}
\usepackage[latin9]{inputenc}
\usepackage{geometry}
\usepackage{afterpage}
\usepackage[round]{natbib}
\usepackage{amsmath}
\usepackage{amsfonts}
\usepackage{amssymb}
\usepackage{amsthm}
\usepackage{lmodern}
\usepackage{rotating}
\usepackage{color}
\usepackage{dcolumn}
\usepackage{setspace}
\usepackage[english]{babel}
\usepackage{eurosym}
\usepackage{esint}
\usepackage{float}
\usepackage{graphicx}
\usepackage{epstopdf}
\usepackage{hyperref}
\usepackage[none]{hyphenat}
\usepackage{booktabs}
\usepackage{threeparttable}
\usepackage{authblk}
\usepackage{pdflscape}
\usepackage{longtable}
\usepackage{multirow}
\usepackage{dpfloat, booktabs}
\usepackage{xfrac}
\usepackage{enumerate}

\newtheorem{assum}{Assumption}
\newtheorem{rem}{Remark}

\newtheorem{alg}{Algorithm}
\providecommand{\keywords}[1]{\textbf{Keywords:} #1}
\begin{document}

\begin{spacing}{1.2}

\title{Fast Algorithms for Quantile Regression with Selection}

\author{Santiago Pereda-Fern\'{a}ndez\thanks {Departamento de Econom\'{i}a, Universidad de Cantabria, Avenida de los Castros, s/n, 39005 Santander, Spain. I would like to thank Manuel Arellano and St\'{e}phane Bonhomme for helpful comments and discussion. This work is part of the I+D+i project Ref. TED2021-131763A-I00 financed by MCIN/AEI/10.13039/501100011033 and by the European Union NextGenerationEU/PRTR. I gratefully acknowledge financial support from the Spanish Ministry of Universities and the European Union-NextGenerationEU (RMZ-18). All remaining errors are my own. I can be reached via email at santiagopereda@gmail.com.}}
\affil{Universidad de Cantabria}
\maketitle

\date

\begin{abstract}
\noindent
This paper addresses computational challenges in estimating Quantile Regression with Selection (QRS). The estimation of the parameters that model self-selection requires the estimation of the entire quantile process several times. Moreover, closed-form expressions of the asymptotic variance are too cumbersome, making the bootstrap more convenient to perform inference. Taking advantage of recent advancements in the estimation of quantile regression, along with some specific characteristics of the QRS estimation problem, I propose streamlined algorithms for the QRS estimator. These algorithms significantly reduce computation time through preprocessing techniques and quantile grid reduction for the estimation of the copula and slope parameters. I show the optimization enhancements with some simulations. Lastly, I show how preprocessing methods can improve the precision of the estimates without sacrificing computational efficiency. Hence, they constitute a practical solutions for estimators with non-differentiable and non-convex criterion functions such as those based on copulas.
\end{abstract}



\end{spacing}

\pagebreak

\input{section1}
\input{section2}
\input{section3}
\input{section4}
\input{section5}
\input{section6}

\bibliographystyle{chicago}
\begin{spacing}{1.0}

\bibliography{C:/Users/santi/Documents/Material/Papers/papers}
\end{spacing}
\pagebreak

\appendix


\end{document}

%% file: section1.tex
\section{Introduction}\label{sec:intro}

Since the seminal work of \cite{Koenker1978}, Quantile Regression (QR) methods have seen significant development in econometrics. However, due to the nature of the optimization problem, its computation has traditionally been demanding. This is particularly relevant when one is interested in the estimation of several quantiles to approximate the entire quantile process or when resampling methods are used for inference. For the standard QR estimator, \cite{Portnoy1997} proposed implementing the estimator using interior point methods, which outperform the simplex methods. Moreover, they proposed a preprocessing algorithm to further improve the speed of computing a single quantile regression. Building on this idea, \cite{Chernozhukov2022} proposed several algorithms to reduce the computational time required to estimate many quantile regressions.

A recent generalization of standard QR is Rotated Quantile Regression (RQR; \citealp{Arellano2017,Arellano2017a}). RQR differs from QR as the quantile index used for the estimation varies by individual, achieved through a copula that rotates the check function used for estimation. This method was proposed to obtain the Quantile Regression with Selection (QRS) estimator, which allows obtaining QR estimates for a model with sample selection on the unobservables. The latter is modeled by a parametric copula, which also needs to be estimated. This estimator has subsequently been adapted to estimate a model with both censoring and sample selection \citep{Chen2022a} or endogenous models with a binary treatment \citep{Pereda2023}.

The goal of this paper is to propose analogous estimation algorithms for QRS, similar to those proposed by \cite{Chernozhukov2022} for QR. However, the estimation of QRS is more complex for two reasons. First, the slope parameters use the conditional copula rather than the quantile index used in QR. Because the former varies for each individual, the algorithms proposed by \cite{Portnoy1997} and \cite{Chernozhukov2022} cannot be directly applied. Second, the estimation of the copula requires the minimization of an objective function that is an integral with respect to the quantile index, using as arguments the slope parameters estimated using RQR for specific values of the copula. This function is non-linear and non-convex, and the integral is approximated using a quantile grid. Therefore, these integrals need to be approximated for a finite number of values of the copula parameter.

In this paper, I propose algorithms to increase the computational speed to obtain RQR and QRS estimates. The algorithm for RQR is an adaptation of Algorithm 2 in \cite{Chernozhukov2022}, which achieves the reduction in computational time through preprocessing. For QRS, I propose two algorithms that combine preprocessing with quantile grid reduction to estimate the copula parameter. The first one obtains the copula parameter with the reduced quantile grid. In contrast, the second one selects a set of candidate values of the copula parameter using the reduced quantile grid and obtains its estimate using the regular quantile grid for these values. While the latter is slower, it results in a more precise estimation of the copula parameter and, consequently, of the slope parameters.

The algorithm proposed by \cite{Chernozhukov2022} takes advantage of the slope parameter estimates at a given quantile as initial estimates at close quantiles. In contrast, here we use the slope parameter estimates at the same quantile, but for a close value of the copula. This is more pertinent, because by reducing the size of the quantile grid, the slope estimates for consecutive quantiles may be more different from each other than the estimates at the same quantile for close values of the copula parameter. This allows for faster estimation of the copula parameter, the slowest step in the estimation of QRS. Then, the estimation for the larger quantile grid is faster, and it can be done by using the slope estimates at consecutive quantiles as initial values.

The asymptotic variance of the QRS estimator is a complex expression that depends on several integrals and density functions. Therefore, it is convenient to carry out inference using resampling methods. Hence, I also propose an algorithm for the multiplicative bootstrap that combines preprocessing with quantile grid reduction to increase computational speed.

Another advantage of the implementation algorithms proposed in this paper pertains to the numerical optimization of the check function. This is usually based on interior-point methods \citep{Portnoy1997} that depend on two optimization parameters. In some cases, the optimization algorithms do not achieve the minimum of the objective function. These suboptimal values do not occur randomly across quantiles or copula parameter values. Consequently, they can add up and lead to a biased estimation of the copula parameters. The algorithms presented in this paper behave to a large extent as a relaxation of the optimization parameters, thus greatly reducing the incidence of this numerical issue without increasing the computational time required.

The rest of the paper is organized as follows: in Section~\ref{sec:model} I present the sample selection model and the estimation procedure. The estimation algorithms are proposed in Section~\ref{sec:algor}, and their performance is shown in Section~\ref{sec:simul}, where I present the results of several simulation exercises. In Section~\ref{sec:num} I also describe some numerical issues regarding the implementation of QR using interior-point algorithms and how the preprocessing implementations presented in this paper can improve the numerical precision of the estimates. Finally, Section~\ref{sec:conc} concludes.

%% file: section2.tex
\section{The Model}\label{sec:model}

The linear quantile model with sample selection proposed by \cite{Arellano2017} can be expressed as follows:
\begin{align}
Y^{*}&=X'\beta\left(U\right)\\
Y&=D\cdot Y^{*}\\
D&=\mathbf{1}\left(V\leq\pi\left(Z\right)\right)
\end{align}
where $Y^{*}$ is the latent outcome, which is observed for participants, \textit{i.e.}, those with $D=1$. In contrast, non-participants have an observed outcome of $Y=0$. The latent outcome is a linear function of several observed covariates denoted by $X$, which are scaled by the slope parameter $\beta\left(\cdot\right)$, which depends on the unobservable $U$. Additionally, the participation decision depends on the unobservable $V$ and the propensity score $\pi\left(\cdot\right)$, which is a function of the vector $Z\equiv\left(Z_{1}',X'\right)'$, consisting on the instrument, $Z_{1}$, and the covariates.

Sample selection is captured by the unobservables $U$ and $V$. Without loss of generality, they are uniformly distributed over the unit interval, and their joint distribution is modeled by a copula, $C\left(u,v;\theta\right)$.\footnote{For expositional simplicity, we consider the case of a copula that does not depend on the covariates. However, this generalization is feasible, as shown in \cite{Arellano2017}.} The lack of selection case corresponds to the independence copula. Additionally, we denote the copula conditional on participation by $G\left(u,v;\theta\right)\equiv\frac{C\left(u,v;\theta\right)}{v}$.

The main restriction used for identification and estimation is as follows:
\begin{align}
\mathbb{P}\left(Y^{*}\leq X'\beta\left(\tau\right)|D=1,Z=z\right)=G\left(\tau,\pi\left(z\right);\rho\right)
\end{align}

This equation relates the conditional distribution of the observed outcome for participants to the amount of selection reflected by the copula. The parameters are identified under the following assumptions \citep{Arellano2017}:
\begin{assum}\label{assum:er}
$\left(U,V\right)$ are jointly statistically independent of $Z_{1}$ given $X=x$.
\end{assum}
\begin{assum}\label{assum:cop}
The bivariate distribution $\left(U,V\right)$, conditional on $X=x$, is absolutely continuous with respect to the Lebesgue measure. Moreover, the marginal distributions of $U$ and $V$ are uniform conditional on all $x$.
\end{assum}
\begin{assum}\label{assum:cont}
$F_{Y|D=1,Z}\left(y|z\right)$, and its inverse in $y$, are strictly increasing. In addition, $C_{x}\left(u,v\right)$ is strictly increasing in $u$.
\end{assum}
\begin{assum}\label{assum:prop}
$\pi\left(z\right)>0$ with probability 1.
\end{assum}
\begin{assum}\label{assum:analy}
Denote the support of $\pi\left(z\right)$ conditional on $X=x$ by $\mathcal{P}_{x}$. $\forall x\in\mathcal{X}$, $\mathcal{P}_{x}\subset\left[0,1\right]$ is an open interval. $\forall \tau\in\left(0,1\right)$, the conditional copula $\pi\mapsto G_{x}\left(\tau,\pi\right)$ is real analytic on unit interval.
\end{assum}

The estimation can be done by the following multi-step procedure:
\begin{enumerate}
\item Estimate the propensity score by $\hat{\pi}\left(Z_{i}\right)\equiv\pi\left(Z_{i};\hat{\gamma}\right)$ (\textit{e.g.}, by MLE).
\item Fix a value of $t\in\Theta$. For $\tau\in\mathcal{T}=\left[\varepsilon,1-\varepsilon\right]$ for some small $\epsilon$, compute $\hat{\beta}\left(\tau;t\right)$ as
\begin{align}\label{eq:betahat}
\hat{\beta}\left(\tau;t\right)\equiv\arg\min_{b\in\mathcal{B}}\sum_{i=1}^{N}D_{i}\rho_{G\left(\tau,\hat{\pi}\left(Z_{i}\right);t\right)}\left(Y_{i}-X_{i}'b\right)
\end{align}
where $\rho_{u}\left(x\right)\equiv xu\mathbf{1}\left(x\geq 0\right)-\left(1-u\right)x\mathbf{1}\left(x<0\right)$ denotes the check function.
\item Estimate the copula parameter by minimizing over $t\in\Theta$:
\begin{align}\label{eq:thetahat}
\hat{\theta}\equiv\arg\min_{t\in\Theta}\left\|\sum_{i=1}^{N}\int_{\varepsilon}^{1-\varepsilon}D_{i}\varphi\left(\tau,Z_{i}\right)\left[\mathbf{1}\left(Y_{i}\leq X_{i}'\hat{\beta}\left(\tau;t\right)\right)-G\left(\tau,\hat{\pi}\left(Z_{i}\right);t\right)\right]d\tau\right\|
\end{align}
where $\varphi\left(\tau,z\right)$ is an instrument function.\footnote{\textit{E.g.}, a polynomial of the propensity score. See \cite{Arellano2017}.}
\item The slope parameters are obtained by $\hat{\beta}\left(\tau\right)\equiv\hat{\beta}\left(\tau;\hat{\theta}\right)$.
\end{enumerate}

Under some additional assumptions, the QRS estimator is asymptotically consistent and converges to a Gaussian process. Moreover, uniform inference can be carried out using resampling methods. See \cite{Pereda2022a} for details.

%% file: section3.tex
\section{Estimation Algorithms}\label{sec:algor}

\subsection{Preprocessing for the Quantile Regression Process}

The QR estimator is obtained through the following problem \citep{Koenker1978}:
\begin{align}\label{eq:check1}
\min_{b\in\mathcal{B}}\sum_{i=1}^{N}\rho_{\tau}\left(Y_{i}-X_{i}'b\right)
\end{align}

As noted by \cite{Portnoy1997}, this minimization problem can be rewritten as:
\begin{align}\label{eq:check2}
\min_{b\in\mathcal{B}}\sum_{i\notin\left(J_{H}\cup J_{L}\right)}\rho_{\tau}\left(Y_{i}-X_{i}'b\right)+\rho_{\tau}\left(Y_{L}-X_{L}'b\right)+\rho_{\tau}\left(Y_{H}-X_{H}'b\right)
\end{align}
where $J_{L}$ and $J_{H}$ are two sets of those observations with respectively negative and positive residuals, $X_{G}=\sum_{i\in J_{G}}X_{i}$ for $G\in\left\{L,H\right\}$, and $Y_{H},Y_{L}$ are two chosen values that are large and small enough, respectively. When these two sets are chosen appropriately \citep{Portnoy1997,Chernozhukov2022}, solving for $\hat{\beta}\left(\tau\right)$ with Equations~\ref{eq:check1} and~\ref{eq:check2} yield the same estimates, but the latter is computationally faster because the effective number of observations is reduced.

This preprocessing idea can be applied to the estimation of RQR. To see this, note that the equivalent of Equation~\ref{eq:check2} can be written as:
\begin{align}\label{eq:check3}
\min_{b\in\mathcal{B}}\sum_{i\notin\left(J_{H}\cup J_{L}\right)}D_{i}\rho_{G\left(\tau,\hat{\pi}\left(Z_{i}\right);t\right)}\left(Y_{i}-X_{i}'b\right)+\rho_{\overline{\tau}}\left(Y_{L}-X_{L}'b\right)+\rho_{\overline{\tau}}\left(Y_{H}-X_{H}'b\right)
\end{align}
where $B_{L}=\frac{1}{\#J_{L}\cdot\overline{\tau}}\sum_{i\in J_{L}}\left(G\left(\tau,\hat{\pi}\left(Z_{i}\right);t\right)-1\right)B_{i}$, $B_{H}=\frac{1}{\#J_{L}\cdot\overline{\tau}}\sum_{i\in J_{L}}G\left(\tau,\hat{\pi}\left(Z_{i}\right);t\right)B_{i}$ for $B=\left\{Y,X\right\}$, and $\overline{\tau}\equiv\frac{1}{N}\sum_{i=1}^{N}G\left(\tau,\hat{\pi}\left(Z_{i}\right);t\right)$. The estimate obtained by solving this minimization problem is numerically equivalent to the estimate given by Equation~\ref{eq:betahat}. However, when $J_{L}$ and $J_{H}$ are appropriately chosen, the effective sample size is reduced similarly to the case of QR.\footnote{The sets $J_{L}$ and $J_{H}$ only include observations from participants by construction.}

Note that the two terms that collapse the observations that are believed to be negative and positive now need to be weighted. This is required because the check function is differently rotated for each individual. Therefore, one can multiply and divide the residuals by $\overline{\tau}$, and then the entire term can be written as a function of $\rho_{\overline{\tau}}$.

\subsection{Rotated Quantile Regression}

Before introducing the estimation algorithms, it is important to discuss the challenges of the QRS estimation procedure. First, Equation~\ref{eq:thetahat} depends on an integral that needs to be approximated. In practice, one can use a grid of quantiles to discretize the estimation, such as the set of percentiles $\tau=\left\{0.01,0.02,...,0.99\right\}$. Second, this equation is non-convex due to the presence of the term $\mathbf{1}\left(Y_{i}\leq X_{i}'\hat{\beta}\left(\tau;t\right)\right)$. Third, the slope parameters (Equation~\ref{eq:betahat}) are computed for the specific value of the copula parameter $\theta=t$, making Equation~\ref{eq:thetahat} non-differentiable. This, together with the second challenge, makes the parameter space $\Theta$ also discretized in practice. This limits the dimensionality of the copula, as well as the possible dependence of the copula with respect to the covariates, as the copula estimation suffers from the curse of dimensionality. Therefore, Equation~\ref{eq:thetahat} is the minimization over a finite set of values. Fourth, the check function in Equation~\ref{eq:betahat} is differently rotated for each individual, preventing us from using the algorithm proposed by \cite{Chernozhukov2022} to estimate the quantile process.

To address the latter point, it is possible to adapt Algorithm 2 in \cite{Chernozhukov2022} to obtain the RQR estimates for the quantile grid $\left\{\tau_{1},...,\tau_{Q}\right\}$ for a given value of the copula parameter $\theta$:\footnote{Matlab codes for this algorithm, and the others presented in this paper are available on the author's website: \url{https://sites.google.com/site/santiagopereda/home/research}.}
\begin{alg}\label{alg:rqr}
Estimate $\hat{\beta}\left(\tau_{1};\theta\right)$ regularly (such as using interior point methods). Then, iteratively for $q=2,...,Q$:
\begin{enumerate}
\item Use $\hat{\beta}\left(\tau_{q-1};\theta\right)$ as a preliminary estimate.
\item Calculate the residuals $\epsilon_{i}=Y_{i}-X_{i}'\hat{\beta}\left(\tau_{q-1};\theta\right)$ and $\zeta_{i}$, a quickly conservative estimate of the standard error of $\epsilon_{i}$. Calculate the $\underline{\tau}-\frac{M}{2n}$ and $\overline{\tau}+\frac{M}{2n}$ quantiles of $\sfrac{\epsilon_{i}}{\zeta_{i}}$, where $\underline{\tau}\equiv\min_{i}G\left(\tau_{q},\hat{\pi}\left(Z_{i}\right);\theta\right)$ and $\overline{\tau}\equiv\max_{i}G\left(\tau_{q},\hat{\pi}\left(Z_{i}\right);\theta\right)$. The observations below this first quantile are included in $J_{L}$; the observations above this second quantile are included in $J_{H}$; the $M=m\left(K\cdot N\right)^{\frac{1}{2}}$ observations between these quantiles are kept for the next step. $m$ is a parameter that can be chosen by the user; by default, it is set to 0.5.
\item Solve the modified problem~\ref{eq:check3} and obtain $\hat{\beta}\left(\tau_{q};\theta\right)$.
\item Check the residual signs of the observations in $J_{L}$ and $J_{H}$:
\begin{enumerate}
\item If no bad signs (or the number of bad signs is below the number of allowed mispredicted signs), $\hat{\beta}\left(\tau_{q};\theta\right)$ is the solution.
\item If less than $0.1\cdot M$ bad signs: take the observations with mispredicted signs out of $J_{L}$ and $J_{H}$ and go back to step 3.
\item If more than $0.1\cdot M$ bad signs: go back to step 2 with a doubled $m$.
\end{enumerate}
\end{enumerate}
\end{alg}

The main difference relative to the algorithm used for the estimation of the standard QR process is in step 2: the quantiles of $\sfrac{\epsilon_{i}}{\zeta_{i}}$ used to determine which observations are included in $J_{L}$ and $J_{H}$ now reflect the rotation of the check function. Moreover, because the amount of rotation varies at the individual level, the minimum and maximum quantiles respectively depend on the minimum and maximum of the conditional copula across individuals. Note that the reason for the algorithm to work is that the preliminary estimate used in step 1 is a good guess when $\tau_{q}$ and $\tau_{q-1}$ are close, as in \cite{Chernozhukov2022}. Overall, this algorithm estimates 1 set of slope parameters using regular methods and $Q-1$ with the preprocessing algorithm.

\begin{rem}
There is no need to initialize the algorithm by estimating the slope parameter at the first quantile of the grid. Instead, one could begin at any other quantile $q'$ (\textit{i.e.}, $q=0.5$), follow Algorithm~\ref{alg:rqr} for $q=q'+1,...,Q$, and modify step 1 of the algorithm by using $\hat{\beta}\left(\tau_{q+1}\right)$ for $q=q'-1,...,1$. This could speed up the estimation, as extreme quantiles tend to be slower to compute.
\end{rem}

\subsection{Quantile Regression with Selection}

The estimation of QRS requires the estimation of RQR for several values of $\theta$. One way to speed up the computation is by using a preliminary, smaller quantile grid to compute Equation~\ref{eq:thetahat} and obtain the estimator $\hat{\theta}$. Then, one can apply Algorithm~\ref{alg:rqr} to compute $\hat{\beta}\left(\tau\right)$ for the larger grid. Using a reduced grid may result in large distances between quantiles, impacting the accuracy of the estimates. This can be overcome by using the estimates from close values of the copula as the preliminary estimates. Define the preliminary quantile grid as $\left\{\tilde{\tau}_{1},...,\tilde{\tau}_{R}\right\}$ and the copula parameter grid as $\left\{\theta_{1},...,\theta_{A}\right\}$. The algorithm is as follows:
\begin{alg}\label{alg:qrs1}
Estimate $\hat{\beta}\left(\tilde{\tau}_{r};\theta_{1}\right)$ for $r=1,...,R$ using Algorithm~\ref{alg:rqr}. Then, iteratively for $a=2,...,A$, $r=1,...,R$:
\begin{enumerate}
\item Use $\hat{\beta}\left(\tilde{\tau}_{r};\theta_{a-1}\right)$ as a preliminary estimate.
\item Calculate the residuals $\epsilon_{i}=Y_{i}-X_{i}'\hat{\beta}\left(\tilde{\tau}_{r};\theta_{a-1}\right)$ and $\zeta_{i}$, a quickly conservative estimate of the standard error of $\epsilon_{i}$. Calculate the $\underline{\tau}-\frac{M}{2N}$ and $\overline{\tau}+\frac{M}{2N}$ quantiles of $\sfrac{\epsilon_{i}}{\zeta_{i}}$, where $\underline{\tau}\equiv\min_{i}G\left(\tilde{\tau}_{r},\hat{\pi}\left(Z_{i}\right);\theta_{a}\right)$ and $\overline{\tau}\equiv\max_{i}G\left(\tilde{\tau}_{r},\hat{\pi}\left(Z_{i}\right);\theta_{a}\right)$. The observations below this first quantile are included in $J_{L}$; the observations above this second quantile are included in $J_{H}$; the $M=m\left(K\cdot N\right)^{\frac{1}{2}}$ observations between these quantiles are kept for the next step. $m$ is a parameter that can be chosen by the user; by default, it is set to 0.5.
\item Solve the modified problem~\ref{eq:check3} and obtain $\hat{\beta}\left(\tilde{\tau}_{r};\theta_{a}\right)$.
\item Check the residual signs of the observations in $J_{L}$ and $J_{H}$:
\begin{enumerate}
\item If no bad signs (or the number of bad signs is below the number of allowed mispredicted signs), $\hat{\beta}\left(\tilde{\tau}_{r};\theta_{a}\right)$ is the solution.
\item If fewer than $0.1\cdot M$ bad signs: take the observations with mispredicted signs out of $J_{L}$ and $J_{H}$ and go back to step 2.
\item If more than $0.1\cdot M$ bad signs: go back to step 2 with a doubled $m$.
\end{enumerate}
\item Obtain $\hat{\theta}$ as the minimizer of Equation~\ref{eq:thetahat} among $\left\{\theta_{1},...,\theta_{A}\right\}$.
\item To obtain $\hat{\beta}\left(\tau\right)$ for $\tau=\left\{\tau_{1},...,\tau_{Q}\right\}$, use Algorithm~\ref{alg:rqr}, where the estimates $\hat{\beta}\left(\tau;\hat{\theta}\right)$ for $\tau=\left\{\tilde{\tau}_{1},...,\tilde{\tau}_{R}\right\}$ have been calculated in step 3 and can be used as the preliminary ones.
\end{enumerate}
\end{alg}

This algorithm speeds up the computation of QRS for two reasons: the preprocessing idea proposed by \cite{Portnoy1997} and the quantile grid reduction used for the estimation of the copula parameter. The latter is possible because one can use the slope parameters for the same quantile and a close copula parameter value as preliminary estimates. Overall, the algorithm estimates 1 set of slope parameters using regular methods, $R-1$ using preprocessing based on near quantiles, $R\left(B-1\right)$ using preprocessing based on near copula parameters, and again $Q$ slope parameters in the final step.


Algorithm~\ref{alg:qrs1} has one disadvantage, as it is not numerically equivalent to the estimate that would be obtained if one used the large quantile grid throughout the entire algorithm. This is because the value of the copula that minimizes Equation~\ref{eq:thetahat} may slightly differ depending on the quantile grid used in the estimation. To address this shortcoming, one could modify steps 5-6 to select a small number of values of $\theta$ for which we can compute Equation~\ref{eq:thetahat} using the large quantile grid. The modified algorithm would be as follows:
\begin{alg}\label{alg:qrs2}
Estimate $\hat{\beta}\left(\tilde{\tau}_{r};\theta_{1}\right)$ for $r=1,...,R$ using Algorithm~\ref{alg:rqr}. Then, iteratively for $a=2,...,A$, $r=1,...,R$:
\begin{enumerate}
\item Use $\hat{\beta}\left(\tilde{\tau}_{r};\theta_{a-1}\right)$ as a preliminary estimate.
\item Calculate the residuals $\epsilon_{i}=Y_{i}-X_{i}'\hat{\beta}\left(\tilde{\tau}_{r};\theta_{a-1}\right)$ and $\zeta_{i}$, a quickly conservative estimate of the standard error of $\epsilon_{i}$. Calculate the $\underline{\tau}-\frac{M}{2N}$ and $\overline{\tau}+\frac{M}{2N}$ quantiles of $\sfrac{\epsilon_{i}}{\zeta_{i}}$, where $\underline{\tau}\equiv\min_{i}G\left(\tilde{\tau}_{r},\hat{\pi}\left(Z_{i}\right);\theta_{a}\right)$ and $\overline{\tau}\equiv\max_{i}G\left(\tilde{\tau}_{r},\hat{\pi}\left(Z_{i}\right);\theta_{a}\right)$. The observations below this first quantile are included in $J_{L}$; the observations above this second quantile are included in $J_{H}$; the $M=m\left(K\cdot N\right)^{\frac{1}{2}}$ observations between these quantiles are kept for the next step. $m$ is a parameter that can be chosen by the user; by default, it is set to 0.5.
\item Solve the modified problem~\ref{eq:check3} and obtain $\hat{\beta}\left(\tilde{\tau}_{r};\theta_{a}\right)$.
\item Check the residual signs of the observations in $J_{L}$ and $J_{H}$:
\begin{enumerate}
\item If no bad signs (or the number of bad signs is below the number of allowed mispredicted signs), $\hat{\beta}\left(\tilde{\tau}_{r};\theta_{a}\right)$ is the solution.
\item If fewer than $0.1\cdot M$ bad signs: take the observations with mispredicted signs out of $J_{L}$ and $J_{H}$ and go back to step 2.
\item If more than $0.1\cdot M$ bad signs: go back to step 2 with a doubled $m$.
\end{enumerate}
\item Select the P values of $\theta_{a}$ such that the value of Equation~\ref{eq:thetahat} is smallest. Define this set as $\left\{\theta_{1},...,\theta_{P}\right\}$.
\item For $p=1,...,P$, use Algorithm~\ref{alg:rqr} to obtain $\hat{\beta}\left(\tau_{q};\theta_{p}\right)$ for $q=1,...,Q$, where the estimates $\hat{\beta}\left(\tau;\hat{\theta}_{p}\right)$ for $\tau=\left\{\tilde{\tau}_{1},...,\tilde{\tau}_{R}\right\}$, $p=1,...,P$, have been calculated in step 3, and they can be used as the preliminary ones.
\item Obtain $\hat{\theta}$ as the minimizer of Equation~\ref{eq:thetahat} among $\left\{\theta_{1},...,\theta_{P}\right\}$; for $q=1,...,Q$, and $\hat{\beta}\left(\tau_{q}\right)=\hat{\beta}\left(\tau_{q};\hat{\theta}\right)$, which was already obtained in the previous step.
\end{enumerate}
\end{alg}

Compared to Algorithm~\ref{alg:qrs1}, the number of estimated slope parameters is increased by $\left(P-1\right)Q$. Despite being slower, this algorithm is more precise: if the estimate of the copula obtained by following \cite{Arellano2017} is one of the $P$ values selected in step 7, then the copula and slope estimates are the same. Regardless, both algorithms provide a large time reduction relative to the standard algorithm.

\begin{rem}
These estimation algorithms still suffer from the curse of dimensionality if the copula depends on multiple parameters. This is particularly relevant when one suspects that the selection into participation differs for individuals with specific characteristics. One way to address this issue is to split the population into a finite number of groups according to their observed characteristics and separately run the regression for each group.
\end{rem}
\begin{rem}
Another potential concern is that the parametric family of the copula is not correctly specified. A class of copulas that can approximate any well-defined copula are Bernstein copulas.\footnote{Despite this desirable property, these copulas are not the most appropriate to model extreme tail behavior, as the copula and its approximand converge to an arbitrary limit at different speeds \citep{Sancetta2004}.} It is possible to estimate the parameters of QRS using these copulas with the algorithm proposed by \cite{Pereda2023}, which combines random search with a sequential increase of the order of the copula.
\end{rem}

\subsection{Bootstrap}

The asymptotic variance of the QRS estimator is quite complex \citep{Arellano2017,Pereda2023}. Therefore, a convenient way to carry out inference is to use resampling methods, such as the weighted bootstrap \citep{Ma2005}. Its validity, established in \cite{Pereda2022a}, requires the following assumption on the bootstrap weights:
\begin{assum}\label{assum:weights}
Let $W_{i}$ be an \textit{iid}\ sample of positive weights, such that $\mathbb{E}\left(W_{i}\right)=1$, $Var\left(W_{i}\right)=\omega_{0}>0$ and is independent of $\left(Y_{i},D_{i},\zeta_{i},Z_{i}'\right)'$ for $i=1,...,N$.
\end{assum}

The bootstrapped coefficients are then computed as
\begin{itemize}
\item For each repetition $j=1,...,J$, draw the weights $W_{i,j}$ for $i=1,...,N$ that satisfy Assumption~\ref{assum:weights}.
\item Estimate the propensity score using the weights for each observation. Denote the estimate by $\hat{\pi}_{j}^{*}\left(Z_{i}\right)\equiv\pi\left(Z_{i},\hat{\gamma}_{j}^{*}\right)$.
\item Estimate the slope and copula parameters by adding the weights to Equations~\ref{eq:betahat}-\ref{eq:thetahat}:
\begin{align}
&\hat{\beta}_{j}^{*}\left(\tau;t\right)\equiv\arg\min_{b\in\mathcal{B}}\sum_{i=1}^{N}W_{i,j}\mathbf{1}D_{i}\rho_{G\left(\tau,\hat{\pi}_{j}^{*}\left(Z_{i}\right);t\right)}\left(Y_{i}-X_{i}'b\right)\label{eq:betahatbt}\\
&\hat{\theta}_{j}^{*}\equiv\arg\min_{t\in\Theta}\left\|\sum_{i=1}^{N}\int_{\varepsilon}^{1-\varepsilon}W_{i,j}D_{i}\varphi\left(\tau,Z_{i}\right)\left[\mathbf{1}\left(Y_{i}\leq X_{i}'\hat{\beta}_{j}^{*}\left(\tau;t\right)\right)-G\left(\tau,\hat{\pi}_{j}^{*}\left(Z_{i}\right);t\right)\right]d\tau\right\|\label{eq:thetahatbt}\\
&\hat{\beta}_{j}^{*}\left(\tau\right)\equiv\hat{\beta}_{j}^{*}\left(\tau;\hat{\theta}_{j}^{*}\right)\nonumber
\end{align}
\end{itemize}

The estimation of the bootstrapped parameters in the last step can also benefit from preprocessing and quantile grid reduction. However, there is no need to obtain a preliminary estimate, as the actual ones are available. The following is the bootstrap analogue of Algorithm~\ref{alg:qrs1}, \textit{i.e.}, it obtains $\hat{\theta}_{j}^{*}$ from the reduced quantile grid:
\begin{alg}\label{alg:boots1}
\begin{enumerate}
Iteratively for each repetition $j=1,...,J$:
\item For $a=1,...,A$, $r=1,...,R$, use $\hat{\beta}\left(\tilde{\tau}_{r};\theta_{a}\right)$ as a preliminary estimate.
\item Calculate the residuals $\epsilon_{i,j}^{*}=Y_{i}-X_{i}'\hat{\beta}\left(\tilde{\tau}_{r};\theta_{a}\right)$ and $\zeta_{i,j}^{*}$, a quickly conservative estimate of the standard error of $\epsilon_{i,j}^{*}$. Calculate the $\underline{\tau}-\frac{M}{2N}$ and $\overline{\tau}+\frac{M}{2N}$ quantiles of $\sfrac{\epsilon_{i,j}^{*}}{\zeta_{i,j}^{*}}$, where $\underline{\tau}\equiv\min_{i}G\left(\tilde{\tau}_{r},\hat{\pi}_{j}^{*}\left(Z_{i}\right);\theta_{a}\right)$ and $\overline{\tau}\equiv\max_{i}G\left(\tilde{\tau}_{r},\hat{\pi}_{j}^{*}\left(Z_{i}\right);\theta_{a}\right)$. The observations below this first quantile are included in $J_{L}$; the observations above this second quantile are included in $J_{H}$; the $M=m\left(K\cdot N\right)^{\frac{1}{2}}$ observations between these quantiles are kept for the next step. $m$ is a parameter that can be chosen by the user; by default, it is set to 1.
\item Solve the modified problem~\ref{eq:check3} and obtain $\hat{\beta}_{j}^{*}\left(\tilde{\tau}_{r};\theta_{a}\right)$.
\item Check the residual signs of the observations in $J_{L}$ and $J_{H}$:
\begin{enumerate}
\item If no bad signs (or the number of bad signs is below the number of allowed mispredicted signs), $\hat{\beta}_{j}^{*}\left(\tilde{\tau}_{r};\theta_{a}\right)$ is the solution.
\item If fewer than $0.1\cdot M$ bad signs: take the observations with mispredicted signs out of $J_{L}$ and $J_{H}$ and go back to step 3.
\item If more than $0.1\cdot M$ bad signs: go back to step 2 with a doubled $m$.
\end{enumerate}
\item Obtain $\hat{\theta}_{j}^{*}$ as the minimizer of Equation~\ref{eq:thetahatbt} among $\left\{\theta_{1},...,\theta_{A}\right\}$.
\item To obtain $\hat{\beta}_{j}^{*}\left(\tau;\hat{\theta}_{j}^{*}\right)$ for $\tau=\left\{\tau_{1},...,\tau_{Q}\right\}$, use Algorithm~\ref{alg:rqr}, where the estimates for $\tau=\left\{\tilde{\tau}_{1},...,\tilde{\tau}_{R}\right\}$ have been calculated in step 4, and they can be used as the preliminary ones.
\end{enumerate}
\end{alg}

Alternatively, one can select a set of candidate values of $\hat{\theta}_{j}^{*}$ from the reduced quantile grid. Then the minimizer of Equation~\ref{eq:thetahatbt} with the regular quantile grid among those candidates would be the bootstrapped copula parameter, similarly to Algorithm~\ref{alg:qrs2} for the actual estimates:
\begin{alg}\label{alg:boots2}
\begin{enumerate}
Iteratively for each repetition $j=1,...,J$:
\item For $a=1,...,A$, $r=1,...,R$, use $\hat{\beta}\left(\tilde{\tau}_{r};\theta_{a}\right)$ as a preliminary estimate.
\item Calculate the residuals $\epsilon_{i,j}^{*}=Y_{i}-X_{i}'\hat{\beta}\left(\tilde{\tau}_{r};\theta_{a}\right)$ and $\zeta_{i,j}^{*}$, a quickly conservative estimate of the standard error of $\epsilon_{i,j}^{*}$. Calculate the $\underline{\tau}-\frac{M}{2N}$ and $\overline{\tau}+\frac{M}{2N}$ quantiles of $\sfrac{\epsilon_{i,j}^{*}}{\zeta_{i,j}^{*}}$, where $\underline{\tau}\equiv\min_{i}G\left(\tilde{\tau}_{r},\hat{\pi}_{j}^{*}\left(Z_{i}\right);\theta_{a}\right)$ and $\overline{\tau}\equiv\max_{i}G\left(\tilde{\tau}_{r},\hat{\pi}_{j}^{*}\left(Z_{i}\right);\theta_{a}\right)$. The observations below this first quantile are included in $J_{L}$; the observations above this second quantile are included in $J_{H}$; the $M=m\left(K\cdot N\right)^{\frac{1}{2}}$ observations between these quantiles are kept for the next step. $m$ is a parameter that can be chosen by the user; by default, it is set to 1.
\item Solve the modified problem~\ref{eq:check3} and obtain $\hat{\beta}_{j}^{*}\left(\tilde{\tau}_{r};\theta_{a}\right)$.
\item Check the residual signs of the observations in $J_{L}$ and $J_{H}$:
\begin{enumerate}
\item If no bad signs (or the number of bad signs is below the number of allowed mispredicted signs), $\hat{\beta}_{j}^{*}\left(\tilde{\tau}_{r};\theta_{a}\right)$ is the solution.
\item If fewer than $0.1\cdot M$ bad signs: take the observations with mispredicted signs out of $J_{L}$ and $J_{H}$ and go back to step 3.
\item If more than $0.1\cdot M$ bad signs: go back to step 2 with a doubled $m$.
\end{enumerate}
\item Select the P values of $\theta_{a}$ such that the value of Equation~\ref{eq:thetahat} is smallest. Define this set as $\left\{\theta_{1},...,\theta_{P}\right\}$.
\item For $p=1,...,P$, use Algorithm~\ref{alg:rqr} to obtain $\hat{\beta}_{j}^{*}\left(\tau_{q};\theta_{p}\right)$ for $q=1,...,Q$, where the estimates $\hat{\beta}_{j}^{*}\left(\tau;\hat{\theta}_{p}\right)$ for $\tau=\left\{\tilde{\tau}_{1},...,\tilde{\tau}_{R}\right\}$, $p=1,...,P$, have been calculated in step 3, and they can be used as the preliminary ones.
\item Obtain $\hat{\theta}_{j}^{*}$ as the minimizer of Equation~\ref{eq:thetahat} among $\left\{\theta_{1},...,\theta_{P}\right\}$; for $q=1,...,Q$, and $\hat{\beta}_{j}^{*}\left(\tau_{q}\right)=\hat{\beta}_{j}^{*}\left(\tau_{q};\hat{\theta}_{j}^{*}\right)$, which was already obtained in the previous step.
\end{enumerate}
\end{alg}

%% file: section4.tex
\section{Simulations}\label{sec:simul}

To show the performance of the different algorithms, we carry out several Monte Carlo simulations, comparing the speed of each algorithm to obtain the estimates. The data generating process is the following:
\begin{align*}
Y^{*}&=X'\beta\left(U\right)\\
Y&=D\cdot Y^{*}\\
D&=\mathbf{1}\left(V\leq\Lambda\left(Z'\gamma\right)\right)\\
\left(U,V\right)&\sim Gaussian\left(0.5\right)
\end{align*}
where the first component of $X$ is the intercept, and the remaining ones are drawn from a $U\left(2,3\right)$ distribution, $Z$ contains $X$ and a $Z_{1}\sim\mathcal{N}\left(0,1\right)$, $\beta\left(U\right)$ equals $\Phi^{-1}\left(U\right)$ for the intercept, and $U\cdot b_{k}$ for the remaining covariates, $\gamma$ equals -1.5 for the intercept, 2 for the instrument, and 0.1$\cdot g_{k}$, where $b_{k}$ and $g_{k}$ are drawn from a $U\left(0,1\right)$ distribution, $\Lambda\left(a\right)=\frac{\exp\left(a\right)}{1+\exp\left(a\right)}$ and $\Phi\left(\cdot\right)$ is the standard normal distribution cdf. The experiment is repeated for several values of the sample size $N$ and the number of covariates $K$. Specifically, $N=\left\{10000,20000\right\}$ and $K=\left\{2,10,20\right\}$. In all cases, the quantile grid used in the estimation is the set of percentiles, whereas for Algorithms~\ref{alg:qrs1}-\ref{alg:qrs2} the preliminary quantile grid is the set of deciles.

The main results are presented in Table~\ref{tab:times}. The baseline algorithm is by far the slowest. For $N=10000$, using Algorithm~\ref{alg:rqr} repeatedly for each value of the copula constitutes a notable improvement, being 5 to 10 times faster. However, the largest gain comes from reducing the quantile grid, as can be seen from Algorithms~\ref{alg:qrs1}-\ref{alg:qrs2}. This gain is most significant when the number of covariates is large: when $K=2$, Algorithm~\ref{alg:qrs1} is almost 50 times faster than the baseline algorithm, and when $K=20$, the speed factor rises to about 100. Moreover, the difference is also more substantial when the sample size increases. For instance, when $N=20000$, the gain from using Algorithm~\ref{alg:qrs1} is close to 125-fold. As expected, Algorithm~\ref{alg:qrs2} is slightly slower, taking on average 1.8 times the time to compute the estimates, regardless of sample size or the number of covariates.

\begin{table}[htbp]
  \centering
  \caption{Computational time of the algorithms}\label{tab:times}
	\begin{threeparttable}
    \begin{tabular}{cccccc}
    \toprule
				N     & K     & Baseline & Algorithm 1  & Algorithm 2 & Algorithm 3 \\
    \midrule
    10000 & 2     & 295.8 & 64.9  & 6.5   & 12.2 \\
    10000 & 10    & 818.4 & 122.8 & 14.3  & 23.9 \\
    10000 & 20    & 1789.7 & 177.0 & 17.8  & 33.2 \\
    20000 & 2     & 682.2 & 129.6 & 13.1  & 24.6 \\
    20000 & 10    & 1702.4 & 229.5 & 23.5  & 44.1 \\
    20000 & 20    & 3742.8 & 299.7 & 30.1  & 56.7 \\
    \bottomrule
    \end{tabular}\begin{tablenotes}[para,flushleft]
\begin{spacing}{1}
{\footnotesize Notes: Baseline corresponds to the regular implementation that estimates each RQR regularly; Algorithm 1 corresponds to the repeated use of Algorithm~\ref{alg:rqr} for each value of the copula; Algorithm 2-3 correspond to Algorithms~\ref{alg:qrs1}-\ref{alg:qrs2}, respectively.}
\end{spacing}
\end{tablenotes}
\end{threeparttable}
\end{table}

The increased time required by Algorithm~\ref{alg:qrs2} has a side benefit, as it is more precise than Algorithm~\ref{alg:qrs1}. This is shown in Table~\ref{tab:msetheta}, which presents the MSE of the copula parameter for each of the three algorithms. Because Algorithm~\ref{alg:qrs1} is based on the smaller quantile grid, it tends to have the largest MSE among these algorithms, although not by much. Regardless, the MSE of all three implementations are almost indistinguishable from each other.

\begin{table}[htbp]
  \centering
  \caption{MSE of $\theta$}\label{tab:msetheta}
	\begin{threeparttable}
    \begin{tabular}{ccccc}
    \toprule
				N     & K     & Algorithm 1  & Algorithm 2 & Algorithm 3 \\
    \midrule
10000 & 2     & 0.00256 & 0.00265 & 0.00254 \\
10000 & 10    & 0.00215 & 0.00222 & 0.00216 \\
10000 & 20    & 0.00736 & 0.00731 & 0.00755 \\
20000 & 2     & 0.00131 & 0.00130 & 0.00131 \\
20000 & 10    & 0.00136 & 0.00137 & 0.00136 \\
20000 & 20    & 0.00563 & 0.00561 & 0.00555 \\
    \bottomrule
    \end{tabular}\begin{tablenotes}[para,flushleft]
\begin{spacing}{1}
{\footnotesize Notes: Algorithm 1 corresponds to the repeated use of Algorithm~\ref{alg:rqr} for each value of the copula; Algorithm 2-3 correspond to Algorithms~\ref{alg:qrs1}-\ref{alg:qrs2}, respectively; scaled by 1000.}
\end{spacing}
\end{tablenotes}
\end{threeparttable}
\end{table}

The results for the bootstrap are quite similar to those for the main estimates. In Table~\ref{tab:times_bt} I show the computation time for $J=500$ repetitions. Algorithm~\ref{alg:boots1} is the fastest, with Algorithm~\ref{alg:boots2} being slightly slower. The speed gain is even more pronounced than that found for the main estimates, as can be seen by comparing the average computation times between the repeated use of Algorithm~\ref{alg:qrs1} and the other two. The reason for this is the use of the actual estimates for preprocessing, thus eliminating the need to obtain a preliminary estimate.

\begin{table}[htbp]
  \centering
  \caption{Computational time of the bootstrap}\label{tab:times_bt}
	\begin{threeparttable}
    \begin{tabular}{ccccc}
    \toprule
    N     & K     & Algorithm 2 & Algorithm 4 & Algorithm 5 \\
    \midrule
10000 & 2     & 68.2  & 7.0   & 12.7 \\
10000 & 10    & 117.5 & 12.1  & 22.9 \\
10000 & 20    & 166.9 & 17.3  & 32.8 \\
20000 & 2     & 129.7 & 13.6  & 23.2 \\
20000 & 10    & 230.2 & 33.1  & 45.7 \\
20000 & 20    & 303.3 & 31.9  & 59.4 \\
    \bottomrule
    \end{tabular}\begin{tablenotes}[para,flushleft]
\begin{spacing}{1}
{\footnotesize Notes: Average across bootstrap repetitions; Algorithm 1 corresponds to the repeated use of Algorithm~\ref{alg:qrs1} for each value of the copula using the estimates to obtain for each bootstrap repetition; Algorithm 4-5 correspond to Algorithms~\ref{alg:boots1}-\ref{alg:boots2}, respectively.}
\end{spacing}
\end{tablenotes}
\end{threeparttable}
\end{table}

As for the accuracy of the different algorithms for the bootstrap, the results are similar to those found for the estimator (Table~\ref{tab:msetheta_bt}): Algorithm~\ref{alg:boots1} is the least precise of the estimates by a small margin, whereas the level of accuracy of the other two algorithms is approximately the same.

\begin{table}[htbp]
  \centering
  \caption{MSE of $\theta$ in the bootstrap}\label{tab:msetheta_bt}
	\begin{threeparttable}
    \begin{tabular}{ccccc}
    \toprule
    N     & K     & Algorithm 2 & Algorithm 4 & Algorithm 5 \\
    \midrule
10000 & 2     & 0.00249 & 0.00273 & 0.00247 \\
10000 & 10    & 0.00237 & 0.00278 & 0.00242 \\
10000 & 20    & 0.00725 & 0.00862 & 0.00713 \\
20000 & 2     & 0.00219 & 0.00225 & 0.00219 \\
20000 & 10    & 0.00149 & 0.00159 & 0.00149 \\
20000 & 20    & 0.00486 & 0.00506 & 0.00482 \\
    \bottomrule
    \end{tabular}\begin{tablenotes}[para,flushleft]
\begin{spacing}{1}
{\footnotesize Notes: Average across bootstrap repetitions; Algorithm 1 corresponds to the repeated use of Algorithm~\ref{alg:qrs1} for each value of the copula using the estimates to obtain for each bootstrap repetition; Algorithm 4-5 correspond to Algorithms~\ref{alg:boots1}-\ref{alg:boots2}, respectively.}
\end{spacing}
\end{tablenotes}
\end{threeparttable}
\end{table}

%% file: section5.tex
\section{Numerical Precision of the Estimates}\label{sec:num}

The QR estimator does not have a closed-form solution. Therefore, obtaining the estimates requires minimizing the objective function using linear programming. \cite{Portnoy1997} proposed an interior-point algorithm to implement this estimator, which is available in several statistical packages.\footnote{This includes Stata, Matlab, R, or C++. See \url{http://www.econ.uiuc.edu/~roger/research/rq/rq.html} for additional information on the implementation of this algorithm for different statistical software.} This algorithm iterates until either a gap is below a certain threshold or the number of iterations reaches a specific number. Consequently, the objective function may only be \emph{approximately} minimized. In such cases, the estimated parameters would be close to, but slightly different from, the values that minimize the objective function. This presents a trade-off between the precision of the estimates and the time required to obtain them.

This numerical issue might be of modest importance when one is interested in estimating a single QR, or even several of them. However, for the estimation of QRS, it becomes more problematic: the estimation of the copula involves using the RQR estimates for a grid of quantiles, so several approximation errors can lead to a biased estimator of the copula, thereby yielding incorrect slope parameter estimates.

The algorithms presented in this paper reduce the time required to obtain the QRS estimates, allowing one to increase the maximum number of iterations of the interior point algorithm and reduce the maximum admissible gap. However, even without modifying these two optimization parameters, the preprocessing algorithms presented in this paper yield estimates that lead to smaller or at least equal values of the objective functions.

To illustrate this, we simulate the same model used in Section~\ref{sec:simul} with a sample size of $n=10000$ and a single covariate $\left(K=1\right)$. The estimators are obtained using three different implementations. First, we use Algorithm~\ref{alg:rqr} for each value of the copula considered with the default optimization parameters: at most 50 iterations, and the gap size is no larger than $10^{-5}$. We refer to this as the preprocessing implementation. Second, we use the regular estimation algorithm with the default optimization parameters. We refer to this algorithm as the restricted implementation. Third, using the regular estimation algorithm, but increasing the maximum number of iterations to 100. This is denoted as the unrestricted implementation.

\begin{figure}[htbp]
\caption{Frequencies of non-minimized rotated check functions}
\includegraphics[width=16.5cm]{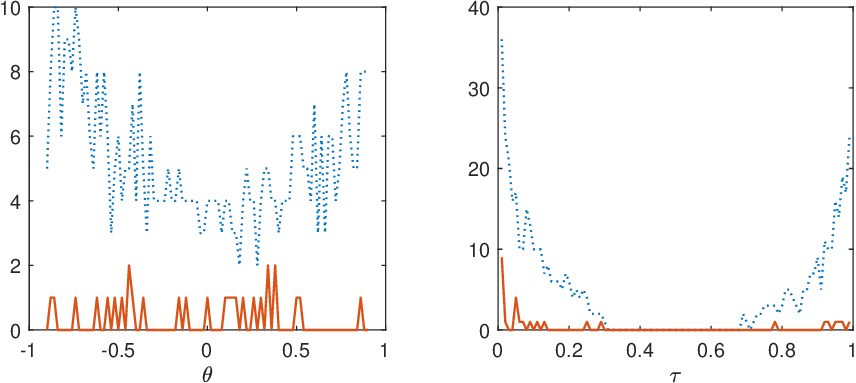}\label{fig:discrep}

{\footnotesize Notes: the left panel denotes the absolute frequencies of values of Equation~\ref{eq:betahat} that are smaller for the estimates from the restricted (dotted blue lines) and unrestricted implementations (solid red lines), than those for the estimates from the preprocessing implementation, for different values of the copula; the right panel denotes the same frequency for different values of the quantile grid.}
\end{figure}

To compare their performance, I examine the value of the rotated check function at the obtained parameter values for each value of the quantile grid (the set of percentiles) and for each value of the copula parameter grid given by $\theta=\left\{-0.90,-0.89,...,0.90\right\}$. Figure~\ref{fig:discrep} shows the number of times the minimized value without preprocessing is strictly larger compared to the value of the preprocessing implementation.\footnote{We consider values to be strictly larger when they are larger than $10^{-3}$, because numerical rounding often causes the objective functions to differ by a tiny amount.} The number of cases in which the preprocessing implementation attains a strictly smaller value is unevenly distributed across quantiles and levels of correlations. In particular, the suboptimal values of the objective function for the other two algorithms occur more often for quantiles that are farther away from the median and higher (in absolute value) levels of correlation. In other words, it tends to occur when the check function is more rotated towards the extremes.

\begin{figure}[htbp]
\caption{Relative minimized rotated check function function}
\includegraphics[width=16.5cm]{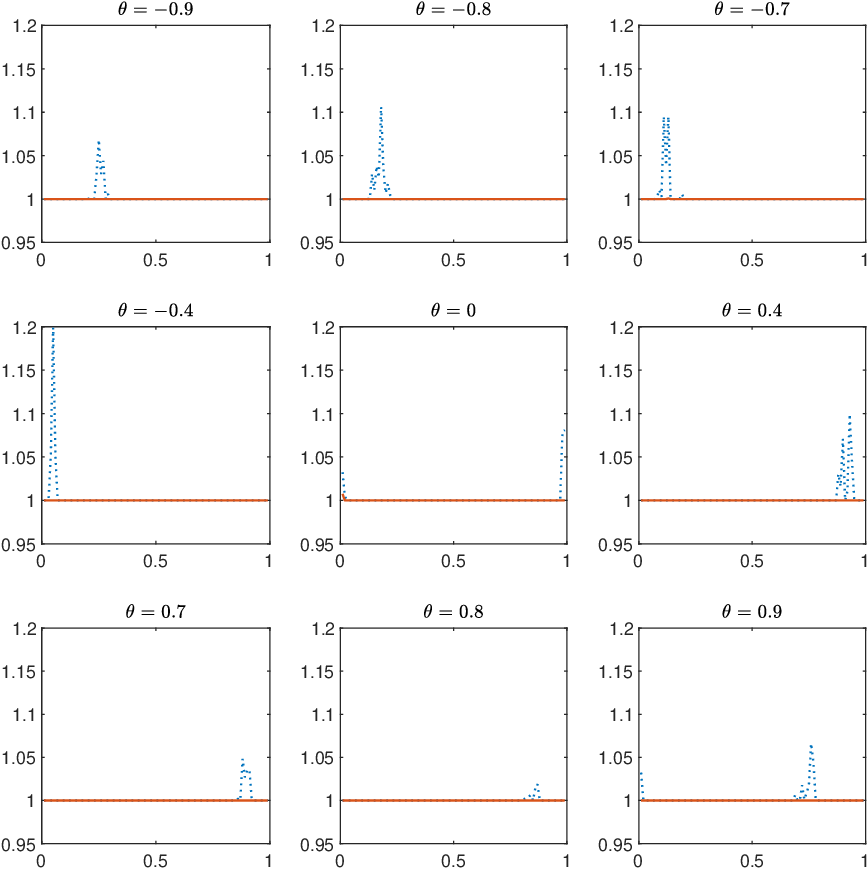}\label{fig:dcheckfn}

{\footnotesize Notes: these figures show, for a number of values of the copula, the relative value of Equation~\ref{eq:betahat} using the estimates from the restricted (dotted blue lines) and unrestricted implementations (solid red lines), relative to that using the estimates from the preprocessing implementation.}
\end{figure}

Moreover, comparing the restricted and unrestricted implementations yields another insightful result: increasing the maximum number of iterations substantially reduces the number of suboptimal values. This reduction occurs both across quantiles and across copula parameter values. In addition, the size of the discrepancy is also reduced in cases where both implementations yield suboptimal results. This can be seen in Figure~\ref{fig:dcheckfn}, which shows the value of the rotated check function for the restricted and unrestricted algorithms, divided by the value for the preprocessing algorithm.


When these lines are equal to 1, the value of the rotated check function is the same for the estimated values obtained with the preprocessing implementation and those with the other two implementations. As expected, this tends to occur for values closer to the median and with modest levels of correlation, and it happens more frequently for the restricted algorithm. However, note also that the line for this implementation is never below the line for the unrestricted one, and in many cases, it is substantially larger, except when they coincide at 1. This means that even when both results are suboptimal, increasing the maximum number of iterations yields values that are much closer to those obtained with the preprocessing algorithm. Additionally, notice that both lines never go below 1, indicating that the value of the rotated check function obtained with the preprocessing implementation is not suboptimal relative to the other two.

This is also reflected in the objective function used to estimate the copula parameter (Equation~\ref{eq:thetahat}). While the function obtained with the unrestricted implementation is almost indistinguishable from the one with the preprocessing implementation, these two are notably different from the function obtained with the restricted implementation. In fact, the value of the copula parameter that minimizes the objective function equals 0.38 using the restricted implementation, it equals 0.40 with the other two implementations.

\begin{figure}[htbp]
\caption{Objective function}
\includegraphics[width=16.5cm]{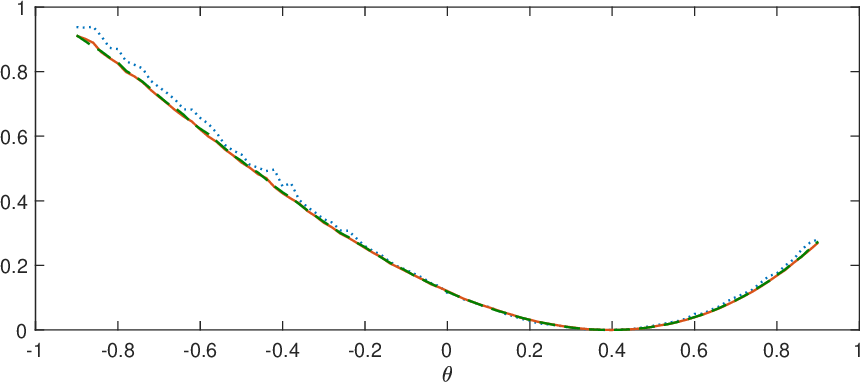}\label{fig:objfn}

{\footnotesize Notes: this figure shows, for a number of values of the copula, the value of the Equation~\ref{eq:thetahat} using the estimates from the preprocessing (dashed green line), restricted (dotted blue line) and unrestricted implementations (solid red line).}
\end{figure}

\begin{figure}[htbp]
\caption{Quantile coefficients}
\includegraphics[width=16.5cm]{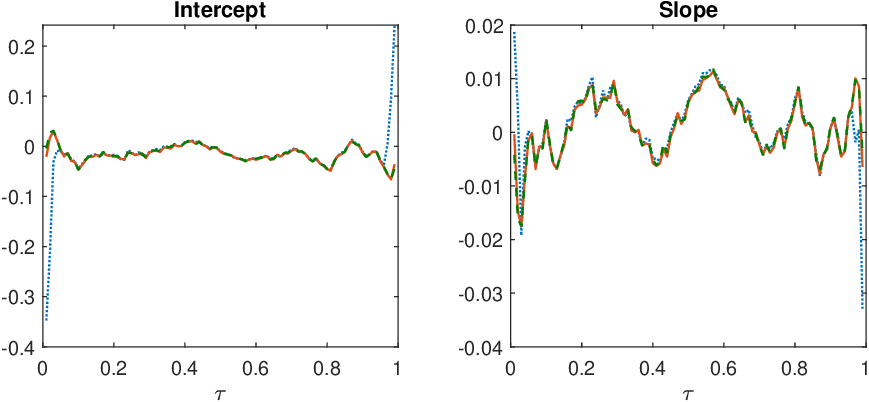}\label{fig:betas}

{\footnotesize Notes: these figures show the difference between the actual value of the quantile coefficients and those estimated by QRS using the preprocessing (dashed green line), restricted (dotted blue line) and unrestricted implementations (solid red line).}
\end{figure}

Unsurprisingly, the estimated coefficients are almost indistinguishable around the center of the distribution (Figure~\ref{fig:betas}). However, for those quantiles closer to the tails, some large differences arise for the restricted implementation relative to the other two.

This highlights a potential caveat of the implementation of quantile regression, that at least can be partially mitigated by the preprocessing algorithms. Hence, for a given level of precision, the time reduction to obtain the QRS estimates using the preprocessing algorithms is even larger than what was reported in Section~\ref{sec:simul}.

%

%% file: section6.tex
\section{Conclusion}\label{sec:conc}

In this paper, I propose estimation algorithms to increase the computational speed to obtain the QRS estimator. These algorithms take advantage of the preprocessing idea proposed by \cite{Portnoy1997} and further developed by \cite{Chernozhukov2022} for standard QR. Even though the minimization problem for QRS is substantially different, by appropriately modifying the estimation algorithm, and also using quantile grid reduction techniques, it is possible to reduce the computational time to obtain the QRS estimator.

We show the performance of these algorithms, verifying that they greatly speed up the computation of the QRS estimator. For moderate sample sizes and numbers of covariates, the computational time is reduced by a factor roughly between 50 and 100. In addition, we show that algorithms based on preprocessing are numerically more precise, which increases the time reduction even more if one considers a similar level of numerical precision.